# Mechanical Manipulation of Graphene Auto-Kirigami with an AFM tip

**Pierce C. Sinnott[1], Majid Fazeli Jadidi[1], Graham L. W. Cross[1]**

**[1] Centre for Research on Adaptive Nanostructures and Nanodevices, Trinity College, Dublin 2, Ireland.**

---

**Abstract**

Graphene auto-kirigami describes the thermodynamically self-driven tearing, sliding and folding of graphene sheets to form micrometre-scale, folded ribbons. However, this process typically requires specialised multi-axial nanoindentation systems or highly laborious AFM-based scratching methods. We here introduce a novel, scalable, wholly AFM-based method to nucleate high yields of ribbons in comparable timeframes to previous multi-axial indentation methods, by AFM-based indentation and "hard tapping", whereby high setpoint AFM imaging can nucleate, manipulate and dynamically image the auto-kirigami ribbons. This can be performed with any conventional AFM, enabling extensional growth, rotation and reversal of ribbons towards potential applications as NEMS devices.

**Introduction**

Folding of graphene and graphite has been explored since Ebbesen et al.'s report of origami-like folding in graphite by STM in 1995[1–3]. A robust, high-yield method of producing folded graphene structures was demonstrated by Annett et al. in 2016. Annett et al. performed nanoindentation with a diamond Berkovich tip on few layer mechanically exfoliated graphene sheets on 300 nm $SiO_2$/Si substrates[4]. Using a 3D nanoindentation system (allowing application of forces in x, y and z axes rather than only in z axis as is typically available in indentation testing) a small amplitude (≈ 1 nm), high frequency (≈ 100 Hz) lateral shearing oscillation was applied during indentation. This resulted in the formation of small, folded tabs of graphene around the indent crater. In many cases, these embryonic tabs can spontaneously grow to micrometre-scale lengths to form folded, tapered ribbons with >70% yield entirely in ambient conditions, as seen in Figure 1 (b) and schematically in (c) and (d). This growth, incorporating tearing, sliding and folding of the graphene sheet, is referred to as auto-kirigami, or AK (where auto refers to the spontaneous, self-driven nature of this growth and *kirigami* refers to the art form of folding and cutting paper to produce complex 3D structures).

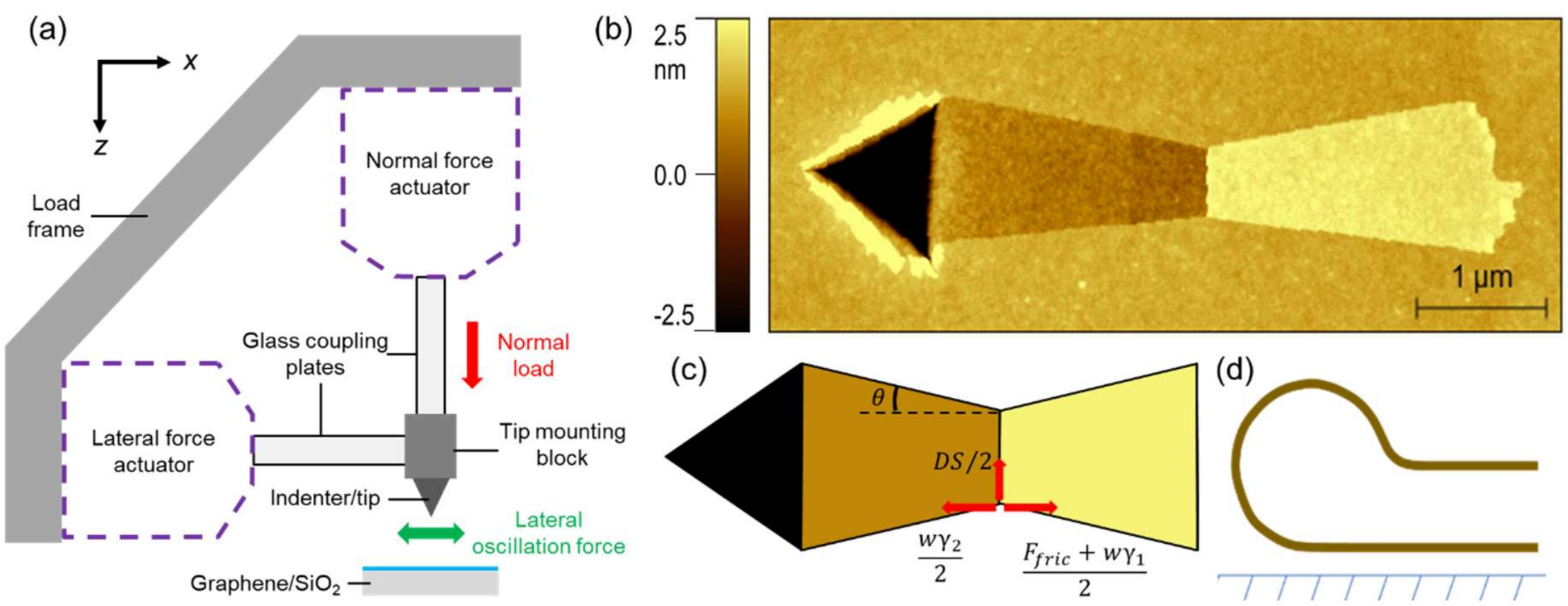

***Figure 1: Method for nucleation of auto-kirigami ribbons.*** *(a) Schematic of Gemini 2-axis isometric indentation system head, capable of sensing and applying forces in two perpendicular axes, as required for SASIN. An analogous system (albeit with 3 axes) was used by Annett et al. in [4]. (b) Example ribbon nucleated from indent produced by SASIN. The graphene peeled, folded and slid spontaneously in ambient conditions, growing to few micrometre lengths. (c) Schematic of AK ribbon, showing advancing forces arising from preferential graphene-graphene adhesion, resisting friction and graphene-substrate adhesion and ribbon tapering arising from a tendency to reduce fold strain energy. (d) Sketch of ribbon fold sideview, showing nanotube-like structure and "hump".*

In this work, we perform equivalent indentation capable of nucleating micrometre scale AK ribbons using a 2-axis indenter system (KLA iMicro nanomechanical tester with Gemini 2D multi-axis transducer), as sketched in Figure 1 (a). We refer to this method as small amplitude sheared indentation nucleation (SASIN).

The AK process is self-driven by an interfacial force which arises as a result of the preferential adhesion of graphene-graphene contact compared to graphene-substrate contact and a tendency to reduce the length of the fold to minimise fold strain energy. This mechanism is supported by atomistic MD simulations[5–7] with qualitatively similar ribbons observed in each (despite a number of differences in the experimental and simulation parameters). Indentation in the absence of lateral oscillations resulted in zero AK ribbons formed, with the application of lateral oscillation therefore considered crucial for successful AK nucleation[4].

Annett et al. described these advancing forces and resisting forces ($F_{advance}$ and $F_{resist}$) acting at each crack tip at a quasi-static, end-of-growth condition by

$$F_{advance} = \frac{\Delta W_{gr-gr} \cdot w\, cos\theta}{2} + D\, S\, sin\theta = \frac{\Delta W_{gr-s} \cdot w\, cos\theta}{2} + \lambda = F_{resist} \quad (1)$$

with graphene-graphene and graphene substrate works of adhesion $\Delta W_{gr-g}$ and $\Delta W_{gr-s}$, fold width $w$, bending stiffness $D$, taper angle θ and rupture energy per unit length λ.

Other methods have also been demonstrated to produce folded AKs. On occasion, folded graphene and tapered AK ribbon structures are observed on graphene after only mechanical exfoliation, as exploited for example by Chen et al. in their study of the graphene fold[8], though this is not a reliable method of production of ribbons. Rode et al. produced twisted bilayer graphene samples on mechanically exfoliated graphene by repetitive contact-mode tracing with a diamond-coated AFM tip at µN loads, producing square micrometre scale areas of TBLG[9]. They ascribe the role of the AFM as simply inducing the initial flip-over to form a fold, with further propagation of the ribbon arising due to the interfacial force mechanism described by Annett et al. This method appears to be low yield however (in agreement with our own testing), with publications using this method reporting 30 and 16 total samples respectively[9,10], in comparison to the tens of ribbons readily produced on a single graphene flake from SASIN.

Sánchez et al. also reported the formation of AK with 1D indentation (with no applied oscillations), alongside conformed, drumhead and broken graphene configurations[11]. Crucially, AK only appear to form with no applied oscillations where the sheet is pre-strained by positioning indents within 1-3 µm of one another (at yields of 0-17%, becoming less likely with larger spacing). An increased yield of up to 49% was observed where vertical oscillations were applied.

Capaldi et al. produced 3D AK structures, with leaflets formed by elastic buckling of up to 100 nm thick flakes of graphite and hBN[12]. These leaflets are either delaminated fully from the substrate (for thinner flakes with thicknesses less than 100 nm) or from remaining graphite layers underneath (for thicker flakes with thicknesses greater than 300 nm). The 3D leaflets protrude from the surface with a plastic hinge rather than being folded “flat” into typically incommensurate contact with the underlying flake as in seen in AK in thinner flakes. These structures were observed following nanoindentation with a diamond spherical tip using the continuous stiffness measurement method with a 2 nm vertical tip oscillation at constant strain rate.

We may simply summarise the above by observing that AK can be produced relatively simply and at high yields by indentation methods compared to the time-intensive, low yield AFM cutting methods, albeit requiring relatively specialised nanoindenter systems capable of applying oscillations in at least one axis. However, we here present a novel method to produce AK with comparatively high yields by AFM-based methods alone, which we dub “hard tapping” (HT). We will then describe a number of related phenomena arising from AFM tip-AK interactions during application of “hard tapping”: rotation of AK, extensional growth of AK and reversal of AK growth. This method also enables nucleation in thick graphitic samples and limited AK nucleation in $MoS_2$ and CVD 2D materials, representing the first instances of nucleation in those materials and indeed the first observed outside of few-layer mechanically exfoliated graphene. We will then propose a “ratcheting” mechanism to describe this AFM tip assisted nanomanipulation of AK ribbons.

*Nucleation*

Indentation is performed using a Bruker Multimode 8 AFM with a DNISP probe (stainless steel cantilever with natural diamond cube corner tip, k = 297 N/m). A maximum load of ≈ 0.5 mN can be achieved with this system, significantly lower than loads of up to 50 mN available using the iMicro dual-axis indentation or 1 N in single-axis indentation. This in turn limits indent depth, with side length of the resulting equilateral triangle-shaped indent impression of no more than ≈ 200 nm. However, an initial ribbon width of ≈ 200 nm is generally below the average required widths of 230±50 nm - 480±60 nm for 1-3 layer graphene for the interfacial force to be of sufficient magnitude to drive ribbon growth per Annett et al.[4]

To overcome this limitation, we consider the geometry of the AFM cantilever during imaging and indentation. AFM cantilevers, when mounted, are oriented with some deflection from the horizontal, often ≈ -12°. The sample is moved by the piezo scanner in the opposite direction to the cantilever during indentation using the aforementioned Bruker indentation mode to prevent ploughing of the surface and to maintain the expected equilateral triangle cube corner indent shape. We instead intentionally under-correct for this ploughing, resulting in pitching of the cantilever with the laser spot moving opposite to the direction of normal deflection and producing deeper indents at higher forces despite equal or less measured cantilever deflection, as in Figure 2 (a).

This is performed using Bruker's nanoindentation software module; while this module allows for automated sample movement during indentation as described above, this is in fact not utilised, with the primary benefits simply being the ability to define arrays of indents to be performed automatically by the system. Comparable indentation which can produce sufficient lengths of torn graphene to nucleate can also be produced purely by ramping the tip, which is available as standard on the Multimode 8 software.

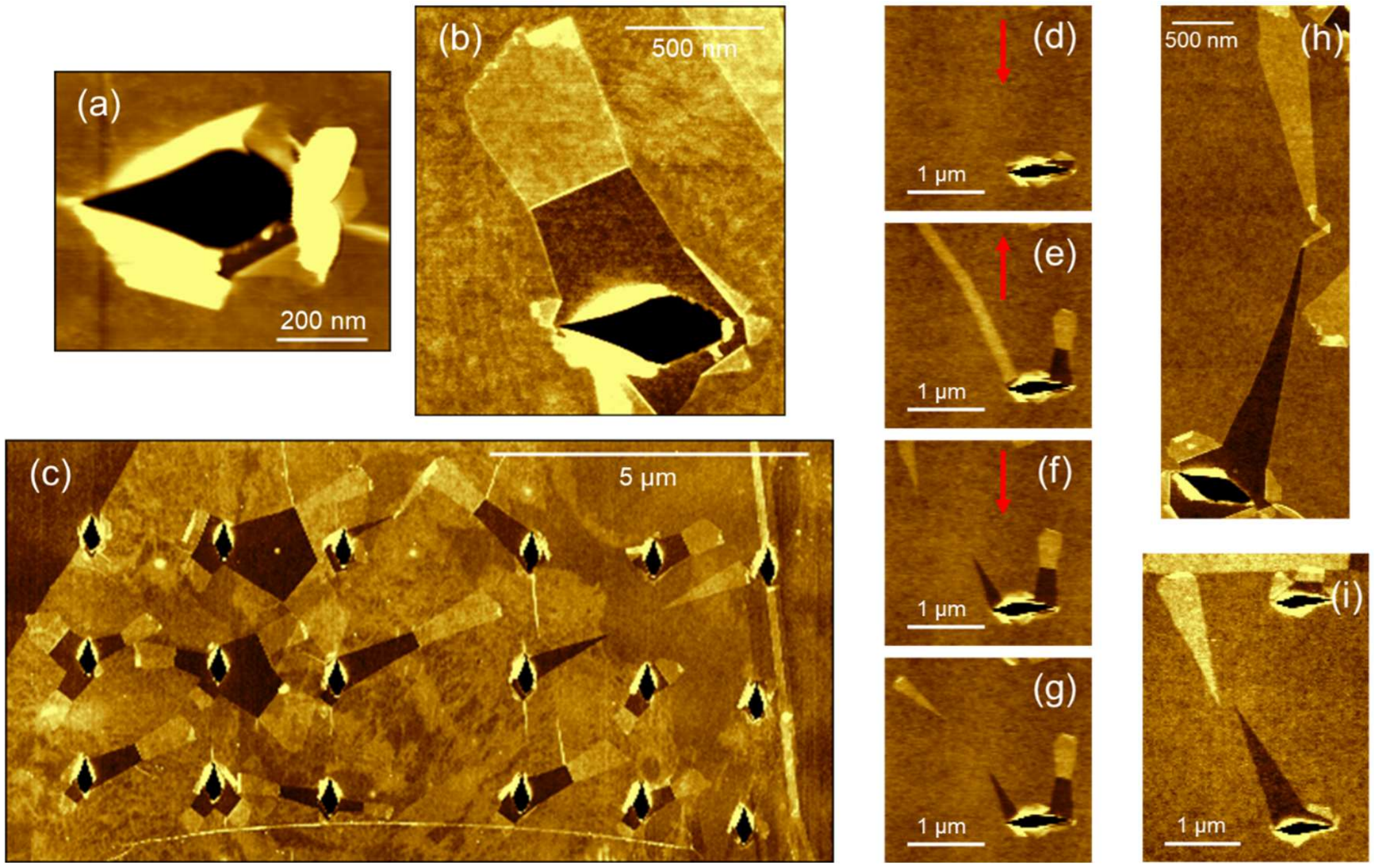

***Figure 2: Ribbon nucleation by hard tapping.*** *(a) A standard indent produced by shearing AFM indentation with a diamond cube corner probe. While some folded graphene is observed around the indent, ribbon nucleation is highly uncommon following only this indentation process. (b) Ribbon nucleation following application of hard tapping on the area around an indent as in (a), with ribbon growth to ≈ 0.5 µm. (c) Widespread, high-yield ribbon nucleation with lengths of ≈ 0.6 – 1.7 µm following hard tapping at 1.4 µN setpoint over 4 scans. This occurs readily even on this relatively contaminated graphene sample. (d)-(g) Sequence of consecutive AFM images (hard tapping applied in (d)-(f) only) showing the dynamic, intermediate ribbon structure during nucleation by hard tapping (e), the ribbon having tapered and fully detached with subsequent transport across the sample (f), and further transport and rotation apparently induced entirely by hard tapping (g). (h) and (i): Further instances of hard tapping resulting in ribbons tapering to near zero width (h) or even fully detaching (i).*

On occasion, indentation by the method described above is sufficient to nucleate ribbons, albeit at relatively low rates compared to other indentation-based nucleation methods, despite the absence of lateral oscillations. We speculate that the ploughing effect of the AFM cantilever enables the formation of these small, folded tabs around the kite-shaped AFM-produced indents as seen in (a). However, we can reliably nucleate ribbons by application of "hard tapping" following AFM indentation, as in Figure 2 (b). Although PeakForce Tapping modes such as PeakForce Quantitative Nanomechanics (PFQNM) are typically selected to operate at few to tens of nN setpoints to minimise damage and wear to both the sample and tip, we instead image using a relatively stiff cantilever (NCL, 48 N/m) with few µN tapping setpoints i.e. as much as 2-3 orders of magnitude greater than is typically applied during tapping mode imaging, hence hard tapping. Upon reaching a sufficiently high setpoint, we can simply and reliably nucleate ribbons not only from a specific indent as in (b) but over multiple indents in scan sizes of tens of micrometres as indicated in (c), nucleating tens of ribbons in minutes.

Upon reaching a threshold setpoint, we observe tip-assisted nucleation of ribbons from indents as in (b) and (c), with further nucleation, growth and rotation of ribbons with increasing setpoint. This can be achieved with setpoints from 500 nN for 1- and 2-layer graphene, often with higher yields for increased setpoint and higher setpoints required for higher layer number samples. Excessively high setpoints (greater than 2-3 µN for few-layer graphene samples) will tend to tear the sheet and any newly formed ribbons however.

Nucleation has also been successful on much thicker graphitic flakes (ranging from ≈ 30 up to ≈ 210 layers) with setpoints between 1-4 µN. We emphasise that in these instances that the ribbon is typically not peeled from the substrate but from remaining underlying graphene layers, with ribbons of varying thickness often observed on a single flake following identical hard tapping e.g. as in Supp. Fig. 3, a 23 nm thick ribbon nucleated from a ≈ 70 nm thick sample, leaving a 47 nm thick peeled area and 93 nm thick stack in the ribbon.

This growth occurs dynamically during imaging- the AFM tip both interacts with the AK ribbon, causing the nucleation, growth, rotation and other phenomena discussed below, while also simultaneously imaging the ribbon responding to the tip. This is highlighted in Figure 2 (d-g). An indent produced by AFM indentation is seen in (d). With application of HT in (e), a long, thin, strip of graphene is seen apparently stretching from the indent to the top of that scan. However, examining the same area in the subsequent scans (f) and (g) instead reveals the tapered, peeled area characteristic of an AK ribbon emerging from the indent, and the ribbon itself detached from the underlying sheet and transported to the upper left corner of the scan. The HT applied in (f) indeed appears to transport the detached ribbon further, to the location observed in (g) where no HT is applied. Other instances of ribbons tapered to near zero width, and wholly detached are also observed in (h) and (i) respectively. We emphasise however that the tapering to near zero final width or complete detachment occurs only at comparatively higher HT setpoints, and that careful selection of setpoint readily allows for large yields of conventional ribbons, as are predominantly observed in (c) for example.

Although hard tapping can also cause damage to the nucleated ribbons and flakes, the larger range of setpoints over which hard tapping can be safely applied and the lesser overall damage observed indicate that this method is highly preferable to contact mode scanning to induce ribbon nucleation or growth. Tens of ribbons can be nucleated from a single array of indents within a single scan at the appropriate setpoint in hard tapping, making this a reliable, high throughput method.

Under Annett et al.'s model, ribbon growth terminates at a fold width characteristic of the graphene sheet thickness, with wider final fold widths for thicker graphene sheets (e.g. 230 ± 50 nm in 1-layer graphene and 480 ± 60 nm in 3-layer graphene). As such, where the initial ribbon width is larger than an expected final fold width, we might expect that the interfacial force is driving ribbon growth and that the tip has perhaps simply induced the necessary initial/embryonic folded tab which can then propagate to form a ribbon "by itself". However, we also perform "hard tapping" on many ribbons which had already reached a final fold width and were stably present in that configuration, having initially been nucleated either by AFM indentation and hard tapping or by SASIN. Applying hard tapping of an appropriate setpoint on such ribbons can induce further growth, indicating that HT does not only induce a "flip-over" that could be analogous to Rode et al.'s AFM-based folding. Instead, HT appears to directly modulate the forces present at the crack tips described in equation (1) or the stress-modified bond breaking process.

Per Annett et al.'s description, fold strain is primarily determined by the shape/curvature of the fold. The shape, which is assumed to remain constant, in turn depends upon the total length of un-adhered graphene in the quasi-static, end of growth conditions. We note that PFQNM imaging of folds shows much higher deformation signals locally along the fold, which may dynamically modify the fold shape and hence the fold strain energy which must be overcome for ribbon growth to occur. It is unclear how HT may affect other parameters described in Annett et al.'s force balance such as graphene-graphene or graphene-substrate adhesion- mechanical loading has been shown to increase the conformity of a graphene sheet to the substrate[13] for example which may modify graphene-substrate adhesion, but the extent to which this effect may contribute during HT is unclear.

While we may also envision a mechanism whereby HT simply overcomes a pinning site or potential barrier to allow the interfacial force to again drive ribbon growth, this interpretation is not supported by experimental observations and ribbon growth appears instead to be driven primarily by the elevated tip-ribbon interactions.

In Figure 3 (a) we observe a ribbon in 7-layer graphene with a length of 0.35 µm, nucleated from a SASIN indent performed several days beforehand. Application of hard tapping in (b) (with slow scan direction from left to right indicated by red arrow) results in extensional growth of the ribbon. This growth is apparently driven entirely/predominantly by the influence of the tip as the interfacial force should no longer be of sufficient magnitude to cause ribbon growth for a ribbon with this fold width in 7-layer graphene. However, in (b), we also reduce the tapping setpoint from 2 µN to only 50 nN during the scan (at the scan line as indicated by red dashed line). Ribbon growth immediately ceases, with the fold and ribbon observed in the successive scan lines of this image. We further note in multiple other examples, where hard tapping is applied over a larger scan area such that multiple indents/ribbons are included, that ribbon growth never continues beyond the edge of the scan area where hard tapping is applied.

When a hard tapping setpoint greater than the minimum necessary to induce nucleation is applied, we also can observe ribbons tapering not only to a new, narrower final width as from (a) to (b) but in fact tapering to a point. In these instances, the ribbon is entirely detached from the underlying flake, except by lying upon it in presumable incommensurate contact, as the ribbon is often transported across the flake, as previously in Figure 2 (d)-(i), in a manner potentially similar to that described by Feng et al.[14] Feng et al.'s work was performed in UHV conditions for islands of only 10s of nm in dimension, with commensuration achieved by thermal transport resulting in translation and/or rotation. Here, transport and rotation appears to be induced wholly by the tip instead. Tapering such that the ribbon detaches or nearly detaches demonstrates that the tip is not merely modulating a kinetic barrier of carbon-carbon bond breaking at the crack tip, or removing a potential pinning site that may have hindered growth by the interfacial mechanism proposed by Annett et al., but instead indicating a strong influence by the AFM tip during hard tapping on this growth.

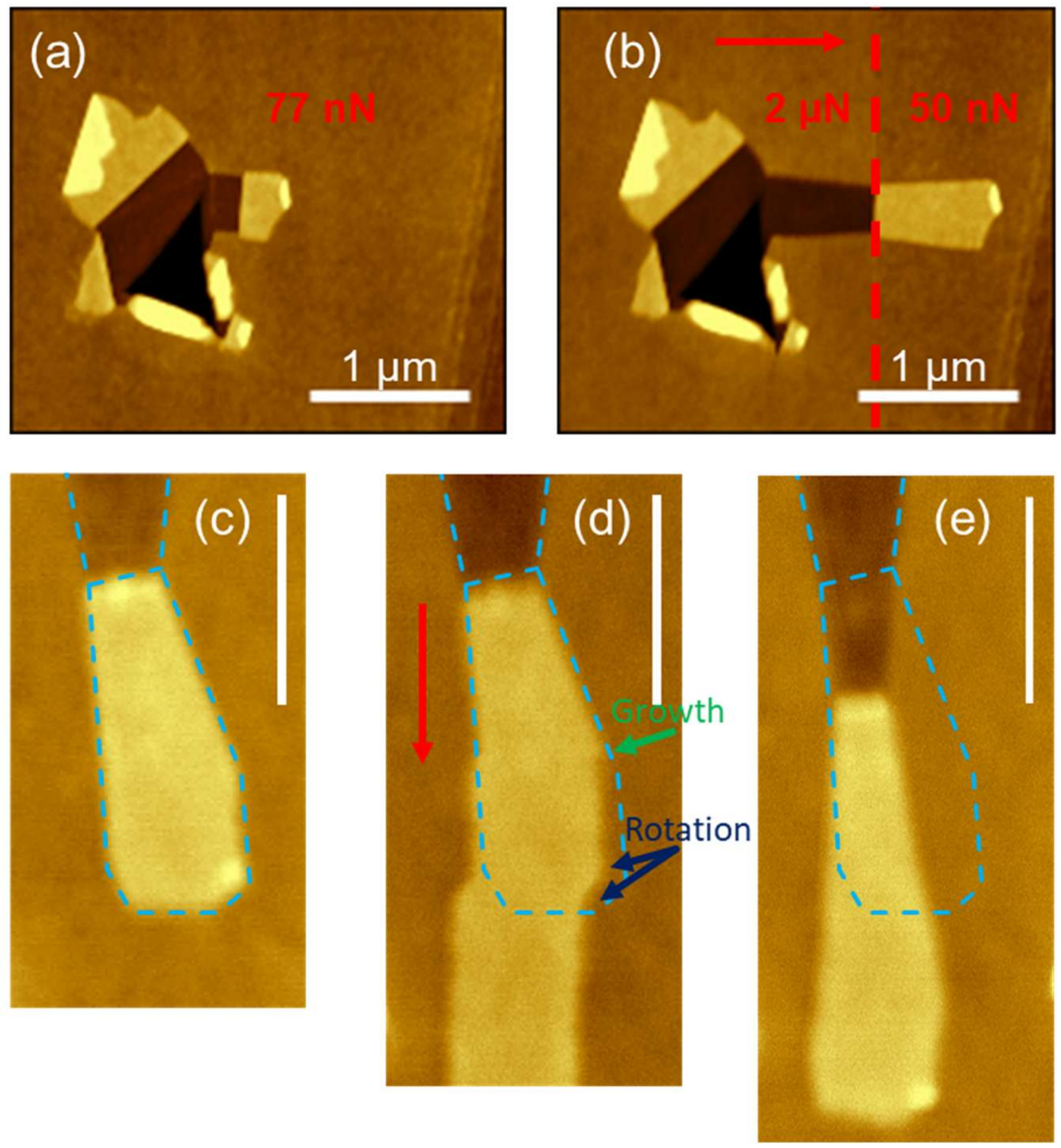

***Figure 3: Ribbon extensional growth due to hard tapping.*** *(a) A ribbon previously nucleated by SASIN imaged by AFM at a moderate setpoint (77 nN). (b) Application of hard tapping at 2,000 nN setpoint results in further extensional growth of the ribbon from (a) in the same direction as the slow scan direction/capture direction, as indicated by red arrow. When this elevated setpoint is reduced to only 50 nN instead during this scan (at the scan line indicated by the red dash line), growth immediately ceases. (c) Ribbon imaged at typical setpoint, with outline indicated by blue dash line. (d) Hard tapping at 1,500 nN setpoint, with slow scan direction indicated by red arrow. Ribbon initially appears unchanged, up to scan line indicated by green arrow where the ribbon no longer follows the previous outline shape and ribbon width now appears constant. The ribbon rotates between the scan lines indicated by navy arrows and continues to grow at apparently constant width in this dynamic image. (e) Ribbon imaged at moderate setpoint after growth and rotation in (d). The ribbon head extends only a short distance from the bottom of the previous hard tapping scan area. Scale bar (c)-(e): 1 μm. Total linear z range: (a) and (b): 7 nm, (c)-(e): 3.5 nm.*

We observe two apparently different mechanisms of the tip interacting with the AK ribbon in Figure 3. In (a) to (b), the imaging reveals the peeled area as the tip moves in the slow scan direction i.e. the ribbon appears to be moving "ahead" of the tip, until the setpoint is reduced and the ribbon is immediately observed. This appears to suggest that ribbon growth occurs due to the tip interactions on or very near to the fold. In contrast, in (c)-(e), we initially see the ribbon in apparently the exact same orientation and position in (d) as it was in (c), up to the scan line indicated by a green arrow. After this scan line, the dynamic ribbon imaging occurring alongside ribbon growth in (d) shows a constant ribbon width for the majority of the ribbon growth, with some rotation of the growing ribbon occurring between the scan lines marked by navy arrows. The latter is by far the most common appearance of the dynamic ribbon growth during hard tapping that we have observed. In this case, as constant ribbon width is observed in (d) but not in the final ribbon shape in (e), it appears that the ribbon growth is occurring at the same speed as the motion of the tip in the slow scan direction.

While we often apply HT in scan sizes encompassing an entire ribbon, or multiple ribbons, HT is also effective at inducing ribbon growth and rotation when applied in smaller areas on a ribbon or on only a small portion of a ribbon. In Figure 4, extensional ribbon growth by HT is successful even where the area of the ribbon is only 0.2% of the total scan area over which HT is applied ((b) - (e) being cropped from scans which originally extended a further 3.5 µm both left and right)). Figure 4 also demonstrates that it is not necessary for HT to be applied directly to the crack tips for extensional growth to occur, as the HT initially is applied to only a small area at the end of the ribbon. Constant ribbon width is again seen in (c) during the HT growth, with tapered edges in (d)-(f).

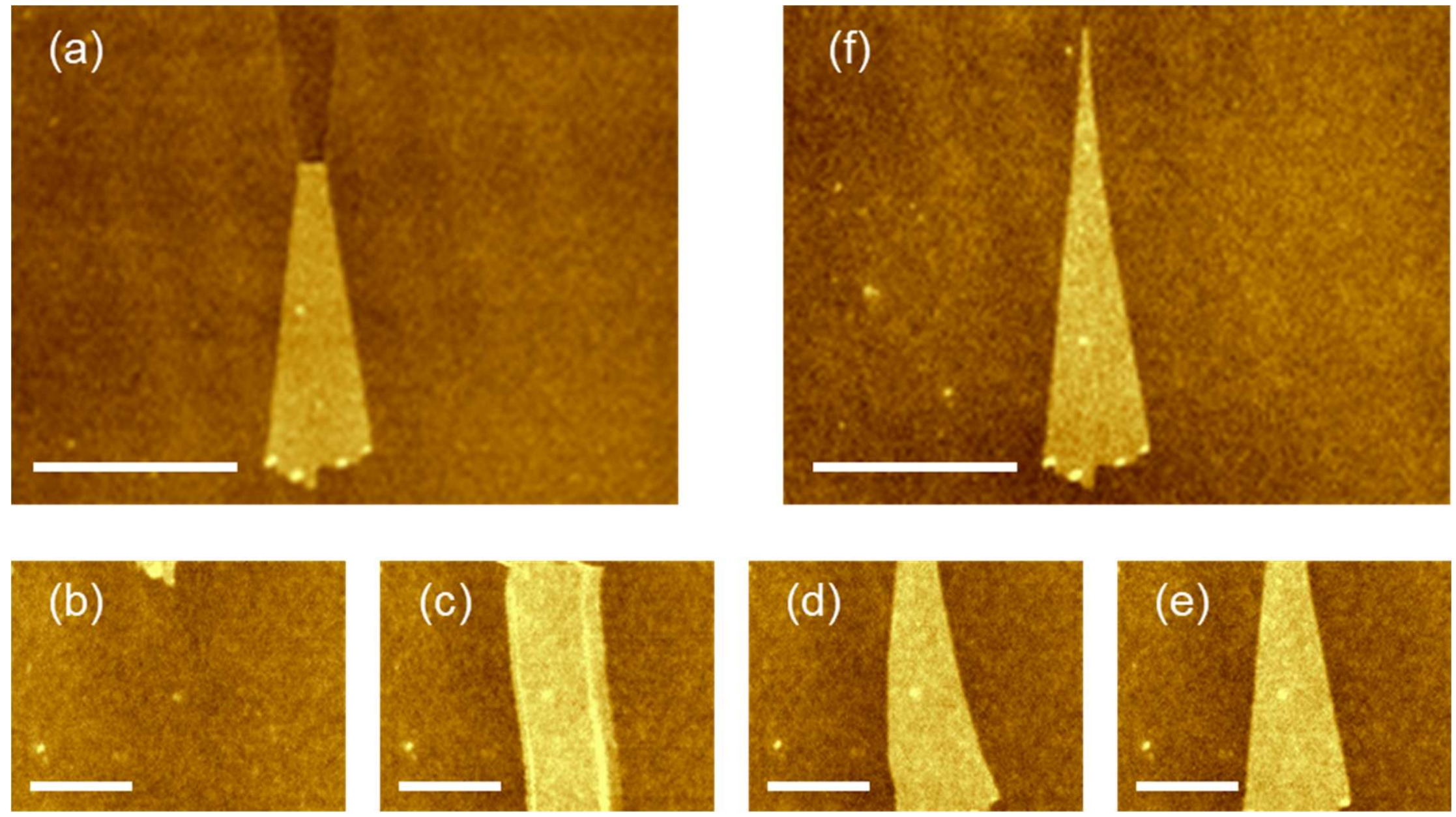

***Figure 4: Ribbon growth by HT applied only to a small area on the ribbon.*** *HT is applied to a ribbon previously nucleated by SASIN (seen before HT in (a), with the second such scan seen in (b) with 1,500 nN setpoint. Extensional ribbon growth occurs in (c), despite the total ribbon area in the scanned location being ≈ 0.05 µm$^2$. HT is also applied in (d), with apparent curved ribbon edges. Straight ribbon edges are instead seen in (e) when HT is no longer applied, as well as in subsequent zoom-out in (f) showing the ribbon tapered to near-zero width. Scale bars: 2 µm in (a) and (f), 1 µm in (b) – (e). Linear z-scale of 2.5 nm in each.*

In addition to extensional growth, on rare occasions reversal of ribbon growth has also been observed as a result of “hard tapping”, as in Supp. Fig. 2, with successive reversal and growth of the ribbon dependent on the alternating slow scan direction. This is currently a rare, stochastic phenomenon which we have observed in only three instances and cannot yet reliably induce. Nonetheless, this represents an important proof-of-concept demonstration of the possibility of triggered ribbon reversal with repeated ribbon peeling and healing. This in turn may enable the use of AK-ribbons as triggered NEMS devices such as valves and actuators as originally postulated by Annett et al.[4]

*Rotation*

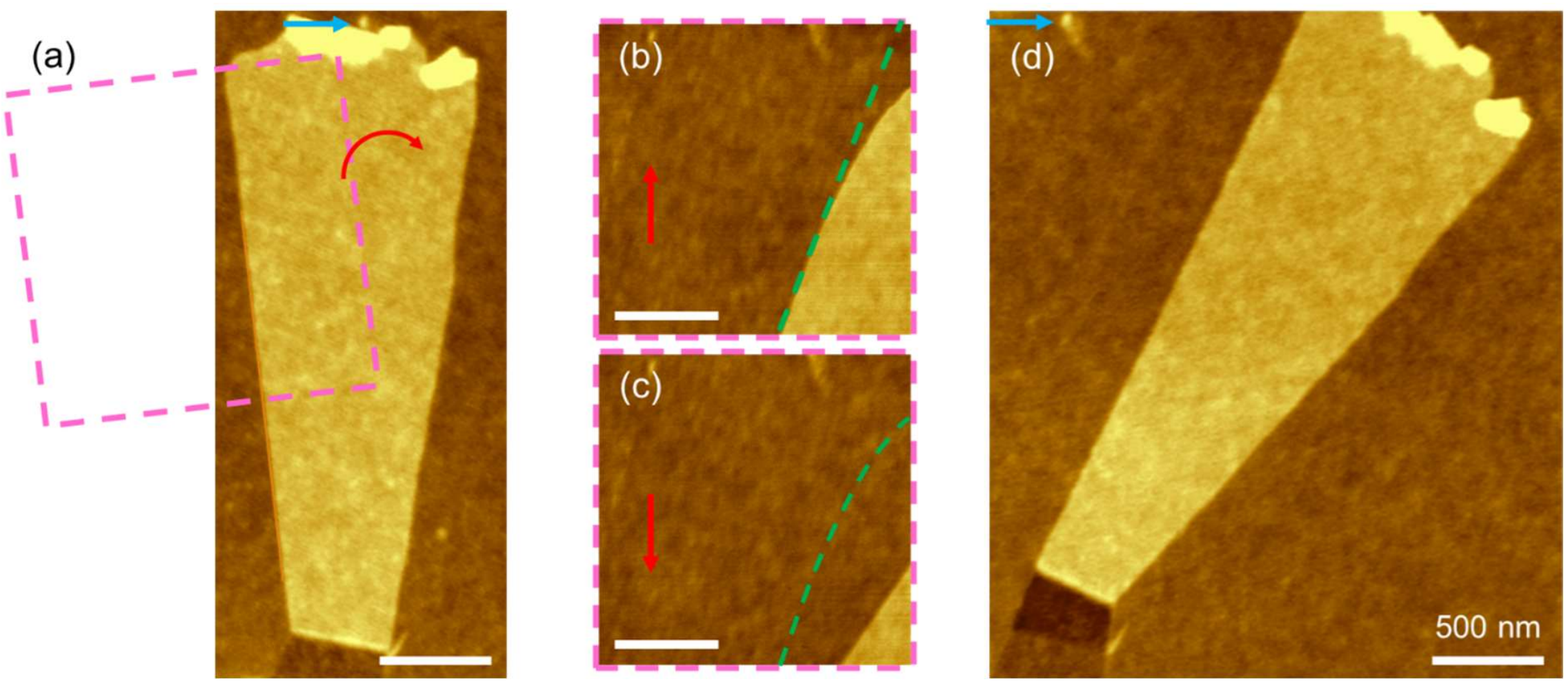

***Figure 5: Precise ribbon rotation by HT on ribbon edge**. A HT scan area is selected such that one ribbon edge will appear vertical (pink dash square in (a)). HT is applied with 1,500 nN setpoint in (b) and (c). Green dash lines indicate the expected ribbon edge based on observed ribbon location in previous scan ((b) and (c) being the 5th and 6th such scan in this location). Slow scan direction indicated by red arrow. (d) Ribbon growth and rotation is observed, as also indicated by relative location of nearby surface feature marked by blue arrow and newly exposed peeled area which appears darker. Total linear z-scale: 3 nm in each. Scale bar: 500 nm in each.*

In a previous work we successfully induced highly localised rotation of single ribbons by contact mode AFM scanning[15], both by scanning over a larger area including the ribbon at a setpoint of 25 nN such that the ribbon rotated in the direction of the fast scan axis into commensuration with the underlying flake, and by scanning over a single line in contact mode at 250 nN setpoint (with this line running across the edge of the rotated ribbon). In both instances, rotation also resulted in other tearing of the graphene flake and notch tearing of the ribbon.

Now, we can achieve precise rotation of ribbons by hard tapping, often with little to no apparent damage to the ribbon itself or to the surrounding graphene sheet. HT is applied to an area including one edge of a ribbon as in Figure 5 (a) (indicated by pink dash outline), with subsequent HT scans in (b) and (c) being the 5th and 6th of a series of 1,500 nN HT scans in this location. Green dash lines in (b) and (c) indicate the expected ribbon edge position copied from the previous scan. Rotation is observed dynamically in (b) as the ribbon edge appears to curve away from the expected straight line. Similarly, copying the ribbon edge position in (b) to (c), and confirming that minimal drift has occurred by comparing nearby surface features such as a dot above the red arrows indicating slow scan direction of (b) and (c), we see that the ribbon has rotated even further to the right. Here, the ribbon has apparently continued growing and rotating even where HT is not applied to the ribbon or ribbon edge, as the ribbon is observed again only in the lower quarter of (c). From (a) to (d) with 6 HT scans, the ribbon rotates 32°, with growth of ≈ 300 nm at the left edge and ≈ 200 nm at the right edge.

*HT mechanism*

From consideration of the above HT related phenomena, we have determined that HT can massively promote nucleation of ribbons from indents produced by AFM, cause extensional growth of existing ribbons which had reached thermodynamic stability, even when applied over small areas of the ribbon far from the crack tips, precisely rotate ribbons and cause ribbon reversal. As ribbon growth and rotation requires fracture propagation at the crack tips, though ribbon growth often occurs when the tip is 100's of nm from the crack tips, we initially suspected that the high setpoint imaging of HT could induce elastic shockwaves in the sheet and/or substrate which, upon propagation to the crack tip, would measurably add an effective driving force perturbation at the activation volume- however, as discussed and further supported experimentally in the supplement, the rapid decay of these stresses with distance makes this highly implausible.

While only retrace channels are captured by default, we can record both trace and retrace height channels to investigate the dynamic effects of the tip on the AK ribbon during HT. In Figure 6 (a), we image an AK ribbon previously nucleated by SASIN at 10 nN setpoint. (b) and (c) show the simultaneously captured trace and retrace images respectively of a subsequent 500 nN setpoint HT scan. Over the first ≈ 700 nm length of the ribbon where HT is applied, no change is observed compared to the initial location in (a). In the trace image (b) the ribbon appears to be rotated to the right, while in retrace (c) the ribbon location appears consistent with only extension of ≈ 100 nm but not rotation. From the ribbon position as imaged immediately afterwards at 10 nN in (d), both extensional growth and rotation have occurred. Similarly in trace and retrace height channels during a subsequent HT scan in (e) and (f), the ribbon primarily rotates to the left-hand side, with a smooth curvature of the ribbon edges observed in (e) and (f) in contrast to the actual straight ribbon edges in (d) and (g).

In any one scan line, the time between the AFM tip's last contact with the ribbon in trace and first contact with the ribbon in retrace is ≈ 0.1 s. Comparing (b) and (c), the ribbon head moves laterally as much as 300 nm in that time.

We propose a ratcheting mechanism, whereby a shear component of the high setpoint AFM tapping is sufficient to laterally "push" the ribbon. From high-speed data capture (HSDC) at 1,000 nN, we determine that the period of a single oscillation is ≈ 0.51 ms, with ≈ 0.11 ms of that in contact and ≈ 0.40 ms out of contact, with a duty cycle of ≈ 0.21. Assuming a linear motion of the piezo tube laterally alongside the 2 kHz modulation of the cantilever, the tip will move laterally as much as 0.2 nm while in contact with the sample during each tap. While tapping imaging modes such as PFQNM minimise lateral forces[16], we suggest that the high normal force applied during HT makes these minimised lateral forces non-negligible in relation to ribbon tearing. With a peak normal force of 1,000 nN, a range of possible graphene friction coefficients of 0.03 – 0.1 results in a peak lateral force of 30-100 nN, applied to a ribbon edge twice in each scan line (to the left edge in trace and right edge in retrace). Annett et al. report an interfacial driving force of approximately 1 nN to drive ribbon growth. We therefore require only a small percentage of these tangential forces to couple to crack-tip driving, with a majority likely dissipated in local slip or translation of the ribbon. This also agrees with our observations that HT growth occurs with greater probability at higher setpoints, as lateral forces would be linearly proportional to setpoint.

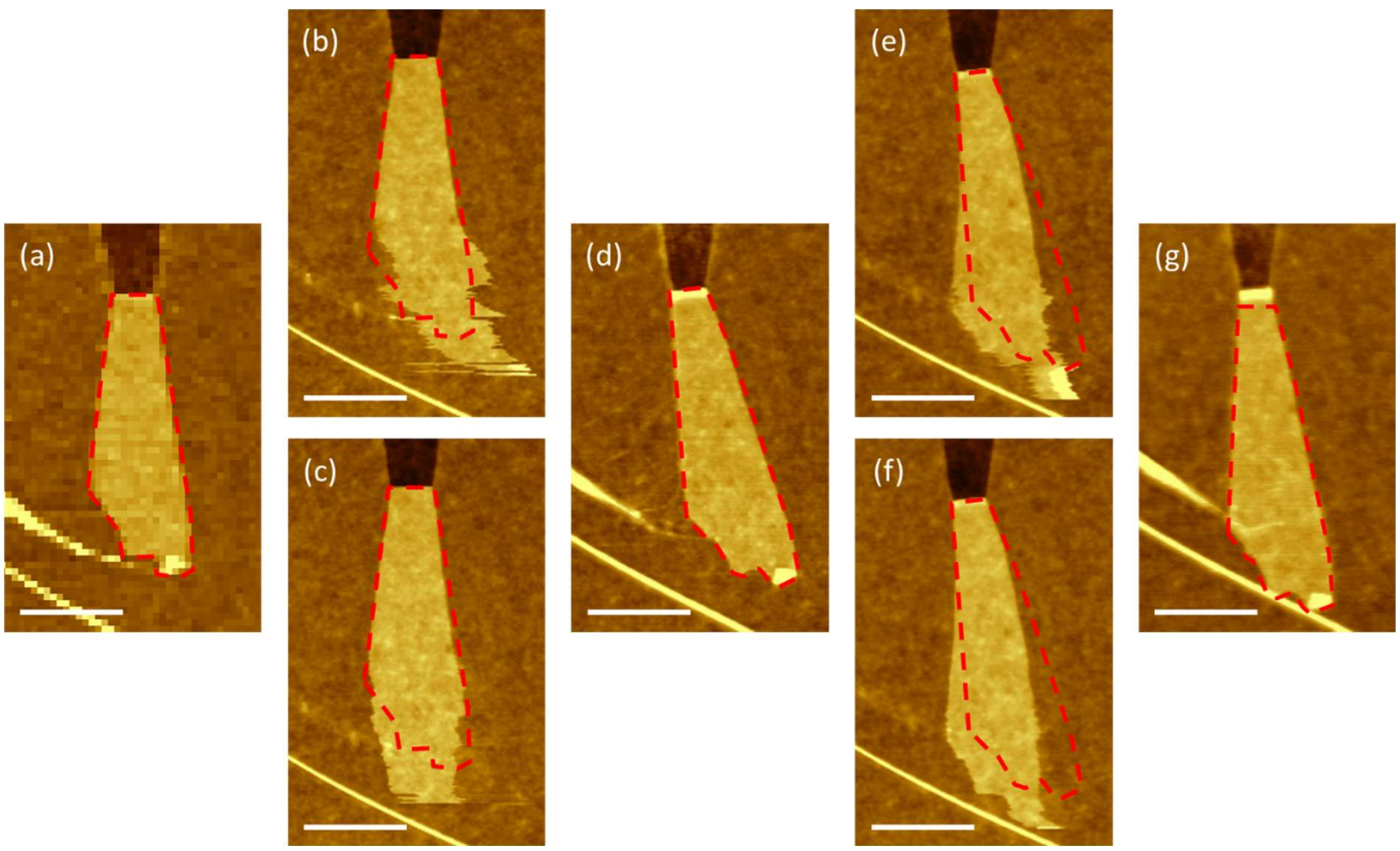

***Figure 6: Simultaneous trace and retrace images during HT.*** *(a) A ribbon previously nucleated by SASIN in 3-layer graphene, imaged at 10 nN setpoint. The setpoint is increased to 500 nN to apply HT in the following image, with the simultaneous trace (b) and retrace (c) height channels of that HT image. Following this single HT scan, the ribbon is imaged again at 10 nN in (d), with similar trace (e) and retrace (f) height channels during a further HT scan at 500 nN setpoint to eventual position imaged at 10 nN in (g).Red dash outlines as a guide to the eye in (b) and (c) are the ribbon outline from (a), and in (e)-(g) are the ribbon outline from (d). Scale bars: 500 nm.*

Ribbon growth can therefore occur with a series of incremental “pushes” at the ribbon edge in each trace and retrace image, corresponding to the different trace and retrace images recorded in a single scan. Similarly, ribbon rotation can occur when the rate of advance at each crack tip is not equal, due to a greater degree of pinning at one edge for example.

This is strongly supported by observations that the rate of ribbon growth typically appears to be approximately the same as the slow scan speed of the AFM, as evidenced by the frequently observed “parallel edges” of the growing ribbon during HT, while the tapered edges are observed both before and after HT. The dependence of ribbon growth rate upon slow scan speeds arises from the shear forces applied to the ribbon edges in each scan line.

Furthermore, even where a ribbon head grows to a position outside of the frame of the AFM scan, the ribbon does not appear to grow to a further distance than that observed in the final scan line. This strongly indicates that the shearing action does not simply remove some pinning site impeding ribbon growth, without which the ribbon's interfacial self-driving force would (entirely by itself) be sufficient to drive further ribbon growth.

## Conclusions

We have introduced HT as a novel, facile method to achieve mechanical manipulation of folded graphene structures. We further demonstrated that HT proves a versatile tool both for production and study of AK ribbons. Nucleation requires only a conventional AFM to readily produce large numbers of AK ribbons without the previous requirement for highly specialised multi-axial nanoindentation systems or extremely limited yields of previous AFM scratching methods for example. The capability of HT to perform AFM-based nanomanipulation with minimal damage to the sample and with predictable outcomes supports potential application of AK ribbons as NEMS devices.

# Supplement: Mechanical Manipulation of Graphene Auto-Kirigami with an AFM tip

*Rotation – with ribbon along fast scan axis*

We typically select the slow scan axis such that it is parallel to the ribbon growth direction, as we observe that HT growth primarily occurs in the slow scan direction. In Supp. Fig. 1, we instead select the fast scan axis as parallel to the ribbon growth direction, with initial ribbon position in (a) before application of hard tapping, dynamic rotation in the direction of the slow scan axis indicated by red arrow in (b), which appears to be reversed in (c) with a curving trace leading to the eventual position of the ribbon head as seen in (d). Hard tapping with setpoints up to 5 µN did not result in further rotation of this ribbon, but instead caused some tearing of the ribbon edges, suggesting that this ribbon may have rotated into commensuration. This rotation appears to be accompanied by a small amount of extensional growth. We suggest

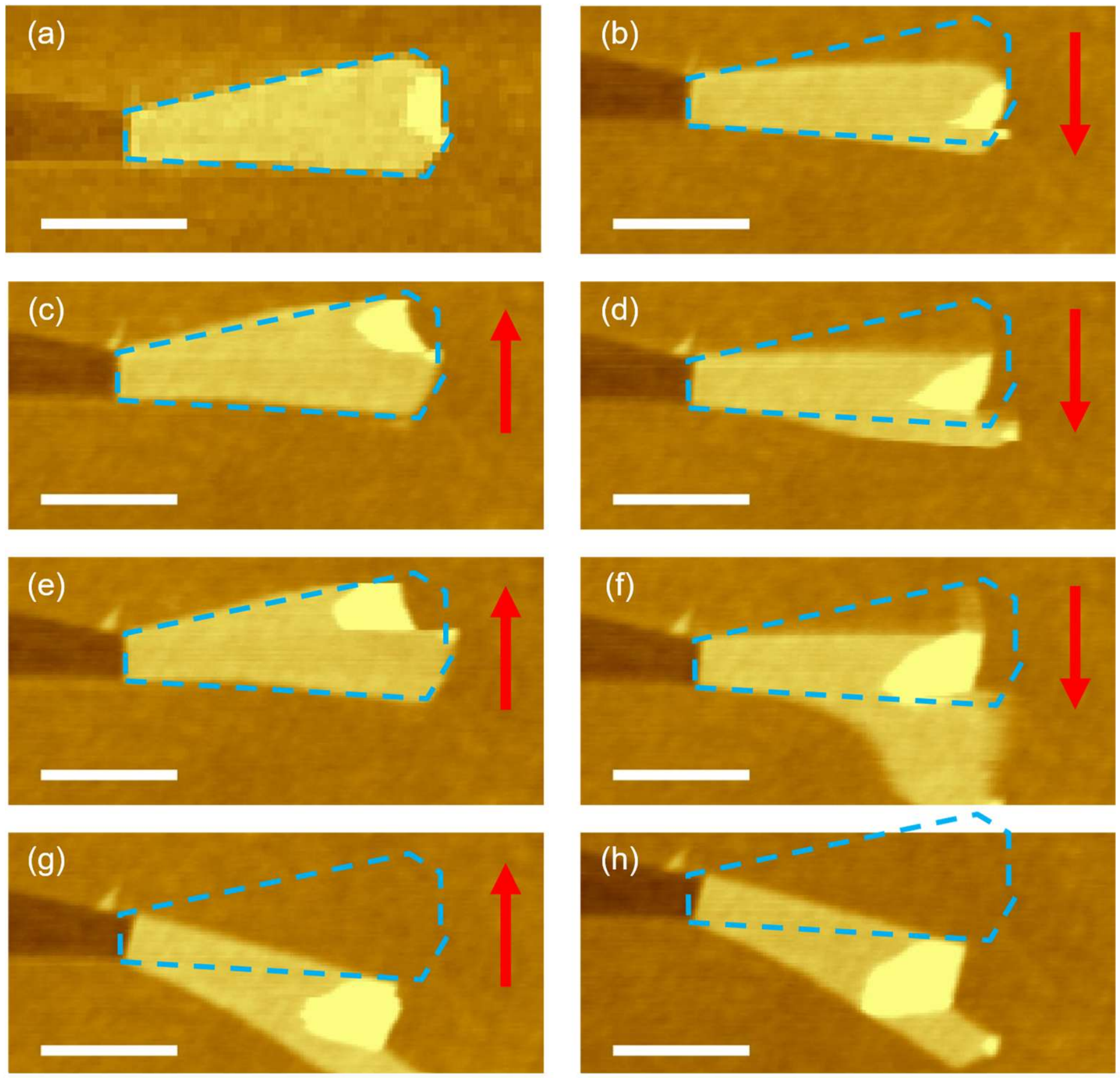

***Supp. Fig. 1: Ribbon rotation with hard tapping**. (a) Ribbon imaged at moderate setpoint in PFQNM, with ribbon outline in (a) highlighted in blue dash lines in (b)-(h). (b)-(g) Consecutive images with hard tapping at 1 µN setpoint. The ribbon rotates in the slow scan direction (indicated by red arrow) in (b), (d) and (f) with dynamic ribbon rotation observed in (f). Hard tapping also results in changes to the shape of the small, folded area of graphene at the ribbon head. Further imaging following (h) at setpoints up to 5 µN did not result in further rotation, suggesting that the ribbon may have rotated into commensuration. Scale bars: 1 µm in each*

that ribbon rotation occurs downwards in Supp. Fig. 1 as the shearing effect of the AFM tip is more effective on the upper ribbon edge which has a greater angular offset from the horizontal (i.e. fast scan) direction, allowing the tip to preferentially "catch" this edge.

We also observe that the height of the fold may increase following rotation, suggesting a small degree of reversal of the ribbon may have occurred (albeit likely on few nm scales), and therefore not typically visible by comparison of the length of the ribbon before and after rotation, as rotation may also be accompanied by some extensional growth (≈ 100 nm total).

*Ribbon reversal*

Supp. Fig. 2 shows a 10 nm thick graphene flake. A PeakForce Tapping setpoint of 3 μN was applied, with no nucleation observed in previous imaging with setpoints increasing in 0.5 μN increments from an initial setpoint of 0.5 μN, with 4 scans at each increment. Supp. Fig. 2 (a) represents the third scan at this setpoint, with (b)-(g) being successive images captured at ≈ 4 minute intervals with alternating scan directions. The slow scan direction is indicated by red arrows in each. The ribbon grows to a length of 4 μm in (b), followed by reversal to a length of 1.6 μm in (c), with similar length of extended and reversed ribbons in (d) and (e). Rotation of the ribbon is also observed in (b) and (d), resulting in an apparently kinked final shape (though we expect that the entire ribbon length has indeed rotated, but that this occurred while imaging near the end of the ribbon so this rotated position is not observed for the majority of the ribbon length).

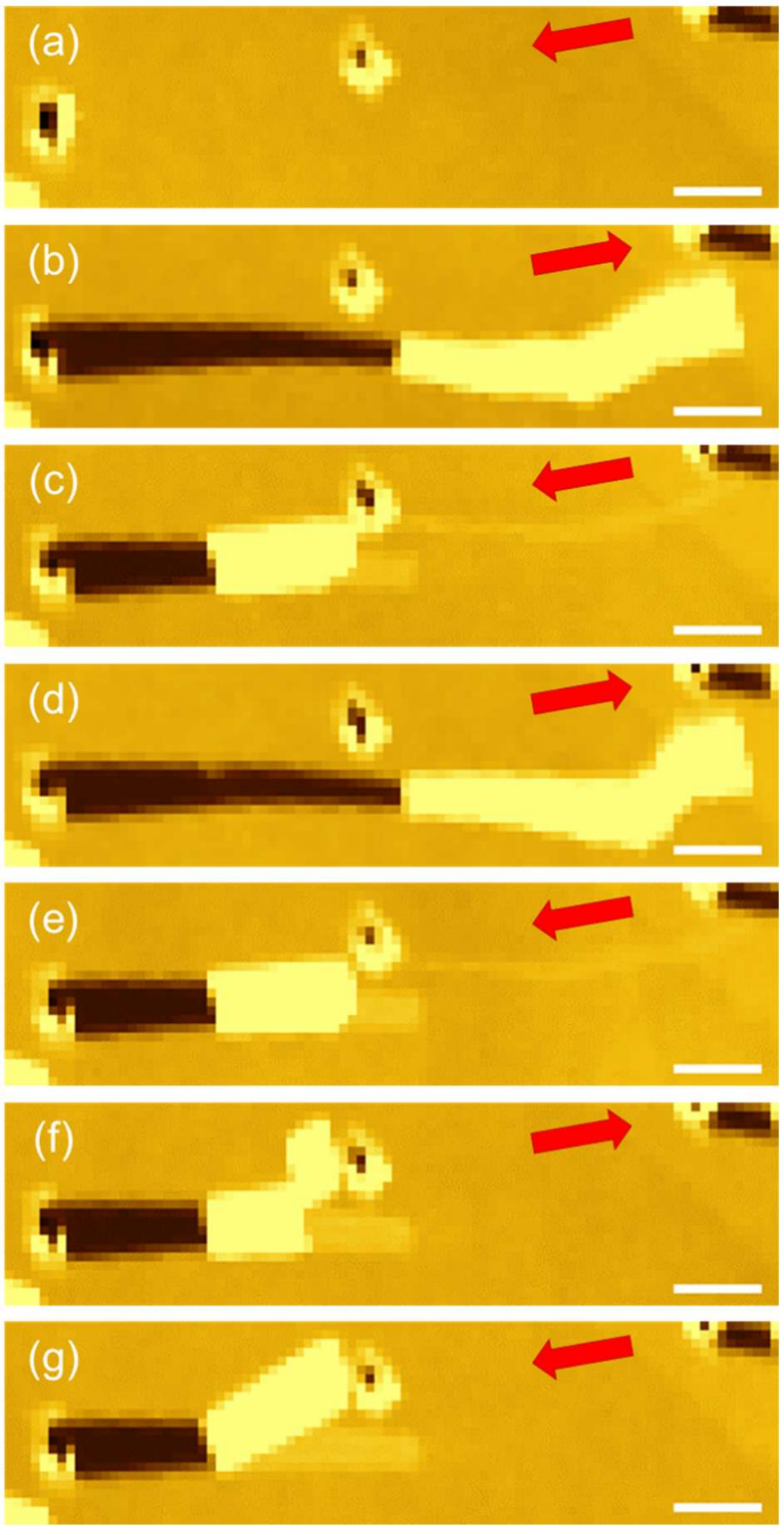

***Supp. Fig. 2: Ribbon reversal induced by hard tapping.*** *3 μN hard tapping setpoint is applied to the consecutive scans shown in (a)-(g). A ribbon is nucleated from a 10 nm thick graphene flake in (b), with the ribbon reversing in (c), subsequently growing again in (d), reversing in (e) before rotating and becoming trapped by a neighbouring indent in (f), preventing further growth. Slow scan direction indicated by red arrow. Scale bars: 500 nm.*

A faint trace is observed in (c) and (e) corresponding to the position of the upper ribbon edge as seen in (b) and (d) respectively. This does not appear to be a permanent defect on the sample surface as it is not seen in (f) and (g), suggesting that this arises from an interaction between the tip and upper ribbon edge during ribbon reversal.

The position of the ribbon fold at the minimum ribbon lengths observed appears consistent across (c), (e), (f) and (g). As the fold is observed in (f) in the same location as in (e), the ribbon does not initially appear to grow during this scan but instead rotates within a single scan line and becomes apparently trapped by the neighbouring indent now located at the rotated ribbon head. No further growth, reversal or rotation was observed even upon increasing setpoint to 5 µN for subsequent scans.

In (c), (e), (f) and (g), a discrepancy is observed in the previously peeled and now apparently healed areas of the ribbons. This peeled-and-rehealed area does not lie smoothly with the surrounding graphene flake, but instead is elevated by 1-4 nm from the surrounding flake. In (b), (d) and (f), rotation of the ribbon appears to occur in the fast scan direction during scanning with slow scan direction along the ribbon length.

We note that, although observed only rarely in this study specifically under the application of hard tapping, this represents the first known instance of reversal of ribbon growth and healing (albeit imperfect) of the torn graphene area. This is an important proof-of-concept demonstration of the possibility of triggered ribbon reversal with repeated ribbon peeling and healing, which may enable the use of AK-ribbons as triggered NEMS devices as originally postulated by Annett et al.[1]

*AK nucleation in graphite, $MoS_2$ and CVD 2D materials*

Previously, Annett et al. reported AK ribbon growth only in 1-4 layer mechanically exfoliated graphene on 300 nm $SiO_2$/Si substrates. More recently, Capaldi et al. reported 3D van der Waals auto-kirigami, with leaflets delaminated from Si/$SiO_2$ produced by nanoindentation of graphite flakes with thicknesses of ≤100 nm (with leaflet thickness equal to flake thickness) extending out-of-plane[2]. Capaldi also reported leaflets with lengths of up to 4 µm and thicknesses on the order of 300 nm that extend out-of-plane from thicker graphite flakes (≈ 1.2 µm). The majority of these remained suspended out-of-plane, with few adhering to the substrate. These 3D structures were formed by plate buckling resulting from in-plane fracture following nanoindentation.

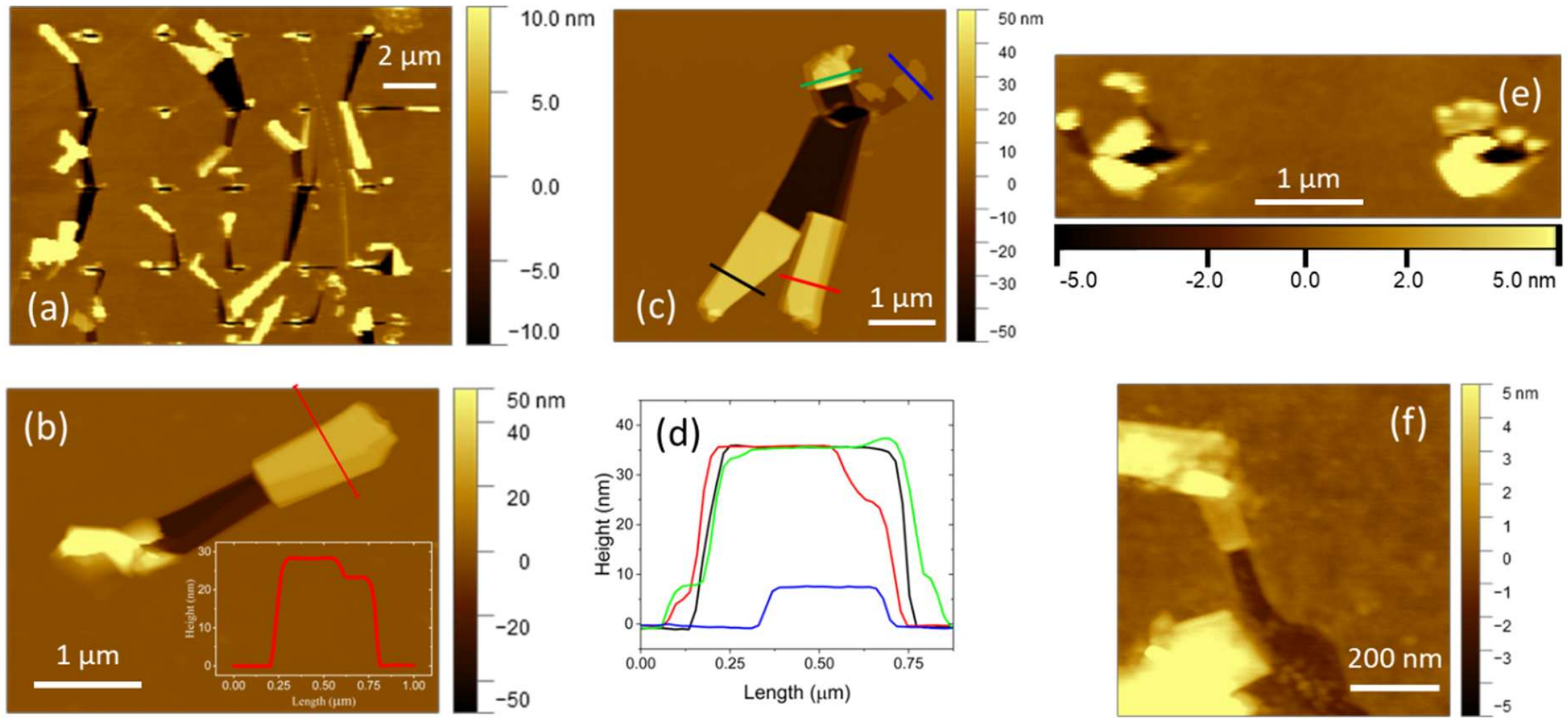

***Supp. Fig. 3: Ribbon nucleation by HT in mechanically exfoliated graphitic sheets and bilayer $MoS_2$.*** *(a) Multiple ribbons nucleated in a 14 nm thick graphitic flake with applied HT of 1 µN. (b) A ribbon nucleated from a ≈ 70 nm thick flake, with varying thickness across the flake (28 nm and 23 nm). (c) While nucleation is observed less commonly in thicker flakes, multiple ribbons can nonetheless nucleate from a single indent following HT. The three thicker ribbons have similar thicknesses of ≈ 36 nm while the fourth (blue line profile in (c) and (d) is 8 nm thick. (e) Limited ribbon nucleation in bilayer $MoS_2$. (f) Zoom-in to ribbon in (e).*

In the fracture mechanics model presented by Annett et al., AK growth requires the 2D material-2D material adhesion energy to be greater than the substrate-2D material adhesion energy. However, ribbon thicknesses in Supp. Fig. 3 (a) vary from 4 nm to the full thickness of the flake of 14 nm, such that the majority of ribbons in this sample are in fact delaminated from graphene rather than from the $SiO_2$ substrate, with no expected preferential adhesion energy to drive the interfacial force described by Annett et al. It instead appears that the energy to drive ribbon growth is provided by the AFM tip itself. AK nucleation is also observed on even thicker flakes, with a single ribbon in (b) having a thickness of 23 nm and 28 nm, delaminated from a ≈ 70 nm thick flake. AK ribbons in graphite as in Supp. Fig. 3 (b) and (c) appear to be fully adhered, unlike the 3D leaflets observed by Capaldi et al.

HT enables AK growth for the first time in non-graphene material, where we have observed very limited ribbon nucleation in few-layer mechanically exfoliated $MoS_2$ (Supp. Fig. 3 (e), (f)). The high setpoints required for ribbon nucleation in this material (5 µN) resulted in significant damage to any newly formed ribbons in the subsequent AFM scan with the same setpoint. HT was also performed on ≈ 175 nm thick mechanically exfoliated $MoS_2$ flake with HT setpoints up to 5 µN, with similar results- while no ribbons were directly observed, the characteristic tapered, peeled area was observed in some instances, suggesting that ribbons were indeed nucleated from these indents during HT, but that the high setpoints destroyed these ribbons.

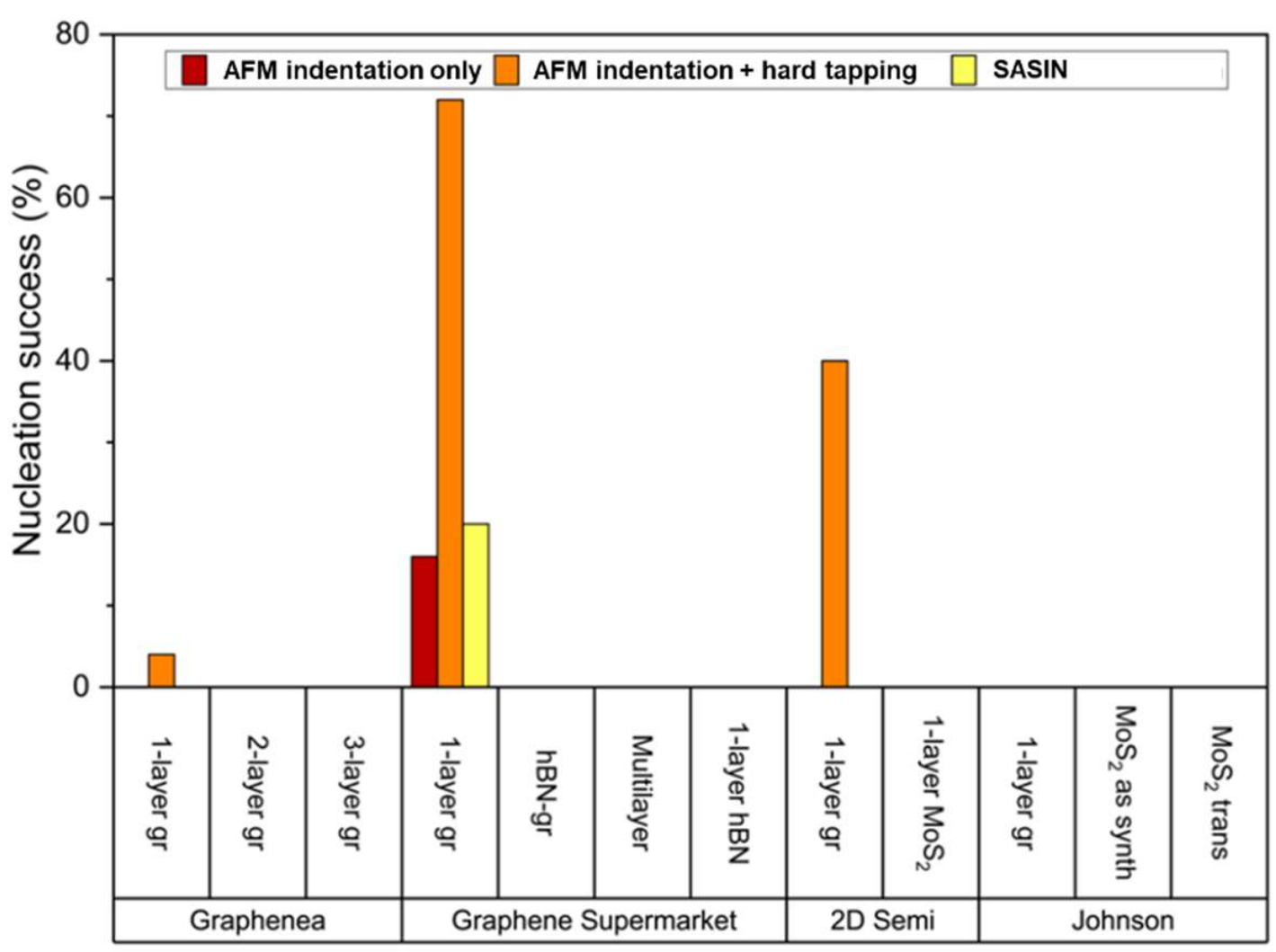

***Supp. Fig. 4: Ribbon nucleation by AFM indentation, AFM indentation followed by HT, and SASIN only on a series of CVD 2D materials from varying suppliers (Graphenea, Graphene Supermarket, 2D Semiconductors, Prof. Charlie Johnson group).*** *While successful nucleation is not observed by any method in most CVD materials, this nonetheless represents the first known AK nucleation in any CVD material. On the most successful sample, HT increases the yield threefold compared to AFM indentation and SASIN methods alone, while HT is required for any AK nucleation to occur on the only other two samples in which AK nucleation is successful.*

HT also enables AK nucleation in CVD 2D materials (Supp. Fig. 4). Three arrays of 25 indents were performed on 12 CVD samples (2 arrays by AFM indentation to 0.5 mN load, and 1 by SASIN indentation using the Gemini dual-axis nanoindenter system to 10 mN load with applied 3 nm lateral oscillation). On one of the 2 AFM indentation arrays, HT was applied in 0.5 µN steps from 0.5 µN to 4 µN, with 4 scans per setpoint i.e. AFM indentation + hard tapping. Each array was surveyed by AFM following indentation, and AFM indentation + hard tapping arrays surveyed again following application of hard tapping. AK nucleation was not observed for a majority of CVD samples, though this nonetheless represents the first known AK nucleation in any CVD material. On the most successful sample (Graphene Supermarket 1-layer graphene), HT increases the yield threefold compared to AFM indentation and SASIN methods alone, while HT is required for any AK nucleation to occur on the only other two samples in which AK nucleation is successful (Graphenea 1-layer graphene and 2D Semiconductors 1-layer graphene). We attribute the failure to nucleate AK ribbons on most CVD samples to a combination of the poorer mechanical quality of CVD sheets compared to mechanically exfoliated flakes, the presence of residual PMMA contamination which was observed on occasion to "gum up" the heads of growing ribbons in CVD samples, and possibly a larger substrate-2D material adhesion energy compared to mechanically exfoliated flakes.

*Elastic shockwaves model*

HT is often applied at setpoints of 500 – 3,000 nN on few-layer graphene to induce ribbon growth. We initially speculated that the high setpoint, high frequency (2 kHz) impacts of the AFM tip could induce an elastic shockwave that would propagate throughout the substrate, resulting in a sufficient barrier change for bond breaking at the crack tip to enable further ribbon growth.

Using AD-42-SS diamond AFM tips, we have also observed that HT at setpoints of 6,000 – 10,000 nN causes plastic deformation of the $SiO_2$. For example, we applied a setpoint of 6 μN in 4 consecutive 100 nm scans centred ≈ 100 nm away from both ribbon crack tips, as in Supp. Fig. 5 (a) and (b) (with 6 μN determined as near the maximum setpoint that could be applied with these tips without causing tearing of the sheet). This setpoint was sufficient to apparently plastically deform the substrate to a depth of ≈ 0.5 nm in the scanned areas without causing visible damage to the sheet visible by high-resolution AFM, as stripe adsorbates are observed continuously both inside and outside the HT scan locations (Supp. Fig. 5 (c)). Despite these high setpoints and close proximity to the crack tips, no ribbon growth was observed, indicating that this barrier change by nearby HT cannot describe the phenomena associated with HT above.

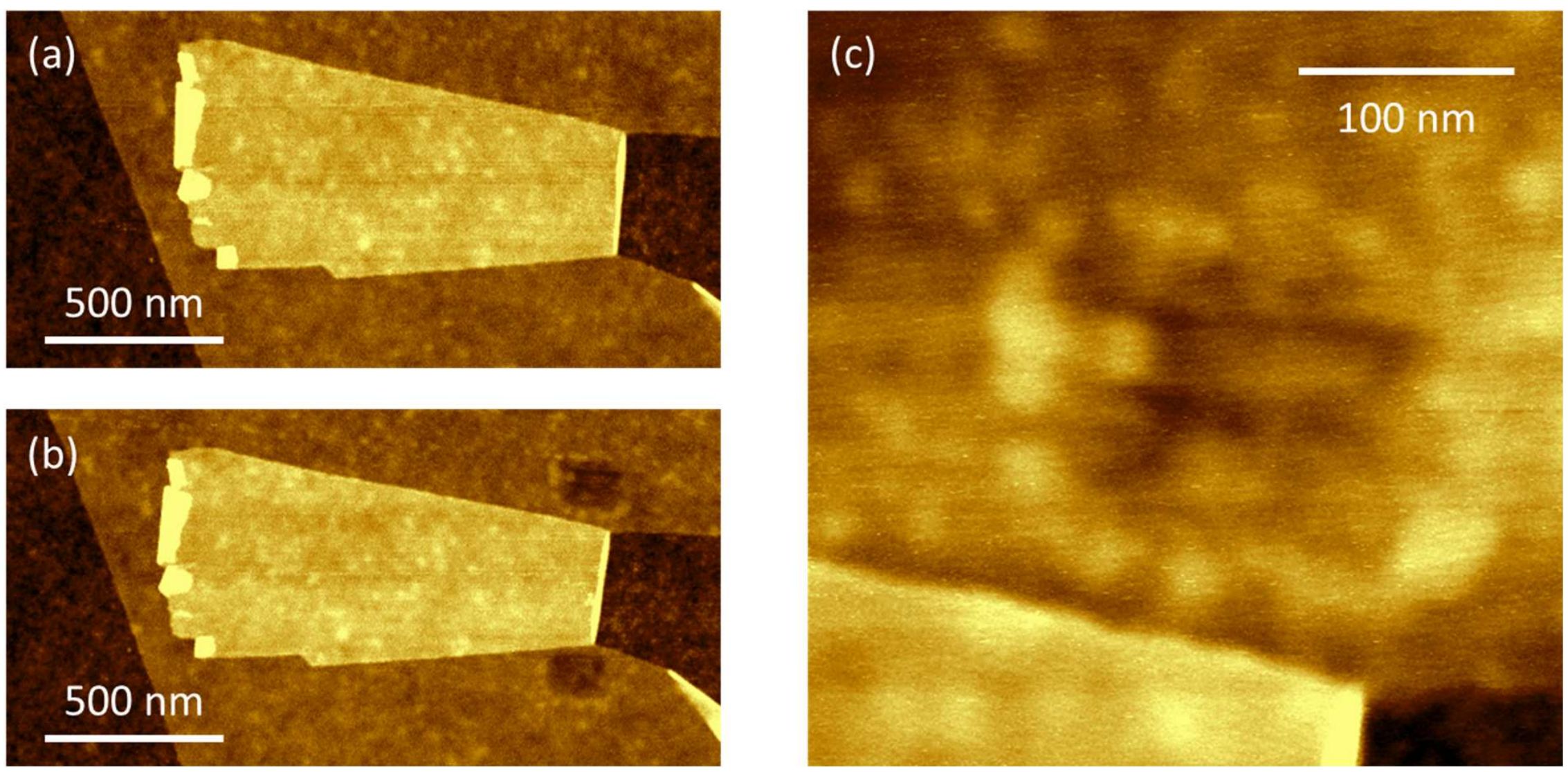

***Supp. Fig. 5: HT at high setpoints near ribbon crack tips.*** *(a) A ribbon previously nucleated by SASIN. 4 100 nm HT scans at 6 μN setpoint are applied, cantered ≈ 100 nm from both the lower and upper crack tip. Impressions from plastic deformation of the substrate following these scans are seen in (b), though these scans were insufficient to result in growth or rotation of the ribbon. The graphene sheet appears undamaged by these HT scans as adsorbate stripes are continuous across the scanned areas, with plastic deformation of ≈ 0.5 nm of the substrate.*

We consider a typical experimental set-up with an AFM scan size of 5 µm, 1 µN setpoint, 0.977 Hz scan rate, at 2 kHz PeakForce frequency, with 512 lines per scan. We will take a tip end radius of 10 nm, typical for a NCL probe as we commonly used (though this probe will become much duller during this high setpoint imaging). Describing the interaction as a Hertzian sphere in contact with an elastic half-space with typical effective modulus $E^*$ of 48 GPa, typical contact radius will be 5-6 nm and peak contact pressure of 10-20 GPa. We note that we cannot directly verify these values by analysis of e.g. the deformation or dissipation channels of PFQNM as these channels become unreliable at highly elevated setpoints.

From high-speed data capture, ≈ 0.11 ms of the ≈ 0.51 ms period of each tap is spent in contact with the sample. The stress field induced by the tip is crucially not a single discrete event but a ≈ 0.11 ms wide pulse, which we will for simplicity consider a triangular ramp-up/ramp-down. For a bilayer graphene sample, we take an activation volume of ≈ 0.015 nm$^3$. In an (overly simple) estimation, we consider that all stored elastic energy in a single tap, determined for a Hertz contact by $U_{el} = \int_0^{u_{max}} F(u)\, du$ as $10^4$ eV, radiates out instantaneously in a wave from a point contact. If we require that the portion of that wave's energy which arrives at the activation volume exceeds $k_BT$, then the tap must be within ≈ 31 nm of the crack tip. This upper-bound estimation is likely much larger than reality, as most elastic energy will instead be recovered.

Annett and Cross describe the fracture propagation necessary for ribbon growth by a stress-modified, thermally activated process with frequency

$$K = v_o \left[\exp\left(\frac{-\Delta F_+}{k_B T}\right) - \exp\left(\frac{-\Delta F_-}{k_B T}\right)\right]$$

with Boltzmann's constant $k_B$, temperature T, natural lattice frequency given by $v_o = {k_B T}/{h}$ and molecular free energies of formation $F_\pm$.

We will consider the stress necessary to lower the activation barrier by, for example, $k_BT$ with $\Delta E \sim \sigma V$, so the necessary stress over the activation volume would be ≈ 0.3 GPa. With Boussinesq scaling of $\sigma \sim {P}/{2\pi R^2}$ [3,4], we obtain an upper bound of R ≈ 23 nm for a 1 µN tap to produce a barrier change of $k_BT$. Only ≈ 70 of the approx. 1 million taps per single scan with the above parameters will take place within this radius. Furthermore, with $\sim {1}/{R^2}$ scaling, we would expect extensional ribbon growth to be much more likely when the tip is closer to the crack tip. Instead, we often note that the tip may travel over half of the length of the ribbon from the crack tip before a change in ribbon position associated with new tearing is observed. As such, we must dismiss this mechanism as the likeliest justification for how HT impacts ribbons.